\documentclass[prl,a4paper,twocolumn,epsfig]{revtex4-1}
\usepackage{epsfig}

\def\be{\begin{equation}}
\def\ee{\end{equation}}
\def\bear{\begin{eqnarray}}
\def\eear{\end{eqnarray}}
\def\bfi{\begin{figure}}
\def\efi{\end{figure}}
\def\btab{\begin{table}}
\def\etab{\end{table}}
\def\bc{\begin{center}}
\def\ec{\end{center}}
\def\bi{\begin{itemize}}
\def\ei{\end{itemize}}

\def\text{\textstyle}

\def\mathswitch#1{\relax\ifmmode#1\else$#1$\fi}

\newcommand\beq{\begin{eqnarray}}
\newcommand\eeq{\end{eqnarray}}

\begin{document}

\title{Evidence of strong higher twist effects in diffractive DIS at HERA at
moderate $Q^2$}

\author{L.\ Motyka}
\email{leszek.motyka@uj.edu.pl}
\author{M.\ Sadzikowski}
\email{mariusz.sadzikowski@uj.edu.pl}
\author{W.\ S\l{}omi\'{n}ski}
\email{wojtek.slominski@uj.edu.pl}
\affiliation{Institute of Physics, Jagiellonian University, Reymonta 4, 30-059
Krak\'{o}w, Poland
}

\begin{abstract}
We study a twist decomposition of diffractive structure functions in the
diffractive deep inelastic scattering (DDIS) at HERA. 
At low $Q^2$ and at large energy the data exhibit a strong excess, up to about 100\%, above the twist-two NLO DGLAP description.
The excess in consistent with higher twist effects.
It is found, that complementing the DGLAP fit by twist 4 and 6 components of the GBW saturation model leads to a good description of data at low $Q^2$.
We conclude that the DDIS at HERA provides the first, strong evidence of higher twist effects in DIS.
\end{abstract}

\maketitle

{\bf 1. Introduction and main results.} The fundamental description of hadron scattering processes is based on the Wilson Operator Product Expansion within Quantum Chromodynamics (QCD). In this procedure the scattering amplitudes are expanded into a series of contributions labelled by twist $\tau$. The twist determines the exponent of leading order scaling behaviour of those contributions in the inverse powers of the characteristic large scale. In Deeply Inelastic Scattering (DIS) the large scale is set by the negative momentum transfer, $Q^2$, from the electron, $e$, to the proton, $p$, corresponding to the virtuality of an exchanged photon, $\gamma^*$. The twist expansion of the DIS cross-section takes the form $\sigma(Q^2) = \sum_{\mathrm\tau} \sigma_{\tau} (Q^2)$, with $\sigma_{\tau}(Q^2) \sim 1/Q^{\tau}$, with $\tau=2,4,\ldots$. In the leading contribution at large $Q^2$, the proton structure enters at $\tau=2$ in the form of universal parton distribution functions (pdfs), whose QCD dependence on the hard scale is described by DGLAP evolution equation. Although the twist-2 description of scattering processes is successful, it has important limitations, that come from neglecting the higher twist (HT) terms.

In a complete and accurate description of hadron scattering the HT contribution should be taken into account. First, they correspond to fundamental hadronic matrix elements that provide additional information on hadron structure, beyond parton densities. Despite the power suppression, the HT effects become relevant below some scale, that depends on the process and the required precision. In particular, a well known problem in precision determination of pdfs is the optimal choice of the lowest scale of fitted data sets. Finally, the HT contributions are related to multiple scattering processes, vital for good description of event properties at hadronic colliders.

Thus far experimental information about HT effects in inclusive hadronic processes is rather poor. A number of phenomenological data analyses were performed for DIS at HERA, showing very weak HT effects for the best measured structure function $F_2$ down to $Q^2 = 1$~GeV$^2$ \cite{SATWIST}. For the longitudinal structure function, $F_L$, however, the HT effect are expected to become relevant below $Q^2 = 10$~GeV$^2$. Unfortunately, present $F_L$ data accuracy at low $Q^2$ is not sufficient for determination of HT effects.      

The diffractive DIS (DDIS) is a semi-inclusive DIS process in which the proton scatters elastically. DDIS events are quite important, as they make up to about 10\% of DIS events at HERA. The standard QCD description of DDIS is based on the leading twist approximation in which the diffractive proton structure functions, $F_2 ^D$ and $F_L ^D$, are expressed in terms of diffractive parton densities.  This approach is justified by Collins factorisation theorem for DDIS \cite{Collins}. However, HT effects are not well understood which may lead to significant uncertainties and contamination in determination of diffractive pdfs. In particular, it is not clear how to choose the lower limit of the scale in DGLAP fits of diffractive pdfs.   

In this paper we perform the first dedicated study of HT effects in diffractive structure functions. We study a spectacular break down of the standard twist-2 DGLAP description of ZEUS data on the DDIS cross-section below $Q^2 = 5$~GeV$^2$ \cite{ZEUS}. In terms of $\chi^2$, the DGLAP fits fail badly below this scale (the relative deviation reaches 100\%) and cannot be extended to the low $Q^2$ region. In the region of low $Q^2$ the data deviations from extrapolated DGLAP predictions grow rapidly with decreasing $Q^2$ and with increasing photon-proton collision energy $W$ \cite{ZEUS}. We find that relative deviations of the data from DGLAP predictions exhibit very steep, power like, $1/Q^2 \div 1/Q^4$ dependence, also the energy dependence is unusually steep. Such behavior of the cross-section is characteristic and natural for HT effects and, by far, too rapid for twist-2 DGLAP evolution. 

In order to estimate the HT contributions we used GBW saturation model~\cite{GBW}. The model is deeply rooted in QCD and offers universal description of variety of observables in DIS, DDIS and vector meson production, see e.g.\ \cite{GBW,KMW,Marquet}. Moreover, the structure of the GBW model allows for a consistent twist expansion \cite{SATWIST}. We decomposed the saturation model predictions for DDIS structure functions into twist components and compared the results to the DGLAP fits and to the data. We found: (i) full consistence of twist-2 component of the GBW prediction with the standard DGLAP prediction --- both unable to describe the low $Q^2$ DDIS data; (ii) improved, but still unsatisfactory description of low-$Q^2$ DDIS data with the complete (all-twists) GBW model; (iii) a good description of the DDIS data with the GBW model truncated to include only its twist-2,~4 and 6 components. This success of the twist-truncated GBW model confirms that the data deviation from twist-2 DGLAP description are consistent with HT effects. We use the twist truncation of GBW to analyse and parameterise the deviation from DGLAP description of the DDIS. In fact, such a truncation may be motivated within QCD, see the last section.   

To summarise the main results of the paper: the available data on low $Q^2$ diffractive DIS show strong break-down of twist-2 DGLAP approximation and the deviations are consistent with sizeable twist-4 and twist-6 effects. Thus, the DDIS data at low $Q^2$ provide the first and strong evidence for HT effects in DIS in the perturbative domain. This opens possibility for further experimental and theory investigations of the HT effects and for significant improvement of the diffractive pdfs.

More details of all the calculations and results will be given in a coming, more extended paper.   \\

\noindent
{\bf 2. The data --- the breakdown of twist-2 description.} 
The DDIS is an $ep$ scattering process 
$e(p_e) \; p(P) \to e(p_e ')p(P')X(P_X)$, with four-momenta $p_e$, $p' _e$ ($P$, $P'$) for the scattering electron (proton). The final hadronic state, $X$, with four-momentum $P_X$, is separated in rapidity from the proton, that scatters elastically. The process is mediated by a virtual photon exchange, $\gamma^*(q)$, with $q = p'_e - p_e$. The DDIS differential cross-section is expressed by invariants: $y = (p_e q)/(p_e P)$, $Q^2 = -q^2$, 
$\xi = (Q^2 + M_X^2) /(W^2 +Q^2)$, $\beta = Q^2 / (Q^2 + M_X^2)$, 
and $t = (P' - P)^2$, where $W^2 = (p + q)^2$ is the invariant mass squared in photon--proton scattering, and $M_X ^2 = P_X^2$ is the invariant mass of the hadronic state $X$. The $t$-integrated $ep$ cross-section reads: 
\be
{d\sigma \over d\beta\, dQ^2\, d\xi} \,=\, {2\pi \alpha_{\mathrm{em}} ^2 
\over \beta Q^4}\, [1 + (1-y)^2] \, \sigma_r ^{D(3)}(\beta,Q^2,\xi)
\ee 
where the ``reduced cross-section'' may be expressed through diffractive 
structure functions, $ F_2 ^{D(3)}$ and $F_L ^{D(3)}$:
$\sigma_r ^{D(3)}(\beta,Q^2,\xi) = F_2 ^{D(3)} - y^2 /[1 + (1-y)^2]F_L ^{D(3)}$,
with $F^{D(3)}_2 = F^{D(3)}_L+F^{D(3)}_T$ and the structure functions $T,L$ 
may be, respectively, expressed through transversaly and longitudinally polarised $\gamma^*$-proton cross sections,
$F^{D(3)}_{L,T}\, =\, (Q^4 / 4\pi^2\alpha_{\mathrm{em}}\beta \xi)\;
d\sigma^{\gamma^\ast p}_{L,T} / dM_X^2$.

In the recent analysis \cite{ZEUS} the ZEUS diffractive data, currently the most accurate in the low $Q^2$ region, were fitted within NLO DGLAP approximation. The data cover the region of $ 2 < Q^2 < 305$~GeV$^2$ but a satisfactory DGLAP description was found only for $Q^2 > Q^2_{\mathrm{min}} = 5\, {\rm GeV}^2$ with $\chi^2 /\mathrm{d.o.f.}$ rapidly growing with decreasing $Q^2_\mathrm{min}$.
Following Ref.\ \cite{ZEUS} we have calculated $\chi^2 / {\rm d.o.f.}$ for subsets of ZEUS LRG data with 
$Q^2 > Q^2_{\mathrm{min}}$,
this time restricted to $\beta > 0.035$ in order to cut off contributions of highly resolved projectiles. 
We have found basically the same behaviour  ---  $\chi^2 /\mathrm{d.o.f.} \simeq 1$ for $Q^2_{\mathrm{min}} > 5$~GeV$^2$, and $\chi^2 / $d.o.f. reaching $\simeq 3$ for the full $Q^2$ range, see the continuous curve in Fig.~1.

\begin{figure}
\centerline{
\includegraphics[width=0.7\columnwidth]{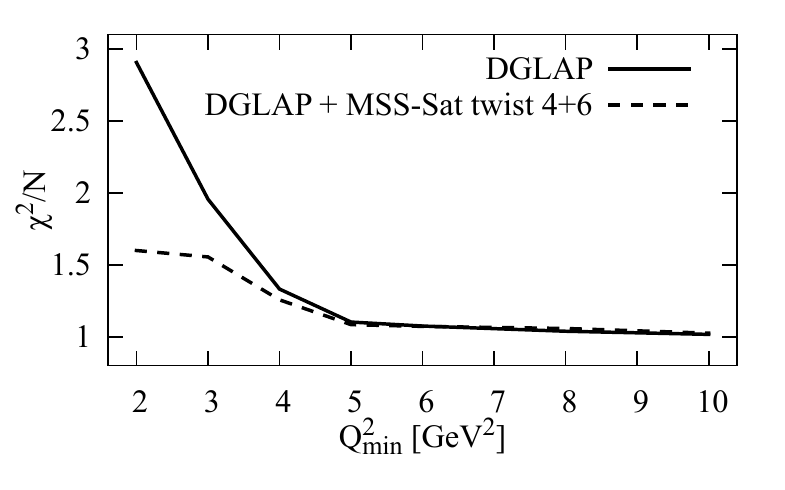}\vspace{-1em}
}
\caption{The $\chi^2/\,{\rm d.o.f.}$ for NLO DGLAP and NLO DGLAP + HT fits to ZEUS LRG data \cite{ZEUS} with $Q^2 < Q^2_{\mathrm{min}}$.
\label{fig1}
}
\end{figure}

In order to analyse the origin of the DGLAP fit deterioration we focus on the low $Q^2$ region.
The ZEUS fits \cite{ZEUS} were performed in the `safe' region, $Q^2>5$~GeV$^2$, and the predictions were extrapolated to lower $Q^2$. The deviations are found to grow rapidly with decreasing $\xi$ and $Q^2$ (see Figs.~3,5 of Ref.~\cite{ZEUS} and Fig.~2). The relative deviation is largest, about 100\%, at the minimal $\xi \simeq 4\cdot 10^{-4} $ and $Q^2 = 2.5$~GeV$^2$. 

We performed a more detailed, numerical analysis, assuming contributions of HTs (the details are presented in the next part of the paper and Fig.~2) and we found that the deviation is consistent with a sum of twist-4 and twist-6 contributions. This, in turn, implies a very strong dependence of the data deviation from twist-2 structure functions, that scales as $1/Q^2 \div 1/Q^4$. However, at twist~2, a possible $Q^2$ modification of the leading $Q^2$~dependence, $\sim\! \mathrm{const}(Q^2)$, due to the DGLAP evolution is moderate $\sim Q^{\gamma}$, with $\gamma \sim \alpha_{\rm s}$. Thus, twist-2 DGLAP evolution is unable to describe the DDIS data below $Q^2 = 5$~GeV$^2$ at low~$\xi$. We shall show that the $Q^2$ and $\xi$ dependencies of the data are, however, consistent with large contributions of HT effects.\\

\noindent
{\bf 3. Twist decomposition of the saturation model.}
In the large energy limit the $\gamma^* p$ scattering may be described in a colour dipole model \cite{DIPOLE,GBW}, where the $\gamma^*$ scattering is factorised into an amplitude of $\gamma^*$ partonic fluctuations and scattering of these states by multiple gluon exchanges. For DDIS one needs to consider the photon fluctuations into a colour singlet $q\bar q$ pair (a colour dipole) and into $q\bar q$-gluon triple, that spans two colour dipoles in the large $N_c$ limit \cite{GBW}. This gives the $t$-integrated $\gamma^*$ cross-sections, 
${d\sigma^{\gamma^\ast p}_{L,T}}/{dM_X^2} = {d\sigma^{q\bar{q}}_{L,T}}/{dM_X^2} +{d\sigma^{q\bar{q}g}_{L,T}}/{dM_X^2}$.
Assuming an exponential $t$-dependence of diffractive cross-sections, one finds for the $q\bar q$ component of $\gamma^*$: 
\beq
\label{cross_sec_qq}
\frac{d\sigma_{L,T}^{q\bar{q}} }{dM_X^2} = 
\frac{1}{16\pi b_D}\int\frac{d^2p}{(2\pi)^2}\int_0^1 dz\, \delta\left(\frac{p^2}{z(1-z)}-M_X^2\right)
\nonumber\\
\times \sum_{\mathrm{spins},\, f}\left| \int d^2r e^{i\vec{p}\cdot\vec{r}}
\psi^f_{h\bar{h},\varepsilon}(Q,z,\vec{r})\sigma (r)\right|^2
\qquad
\eeq
where $b_D$ is a diffractive slope and the sum over spins runs over massless (anti)quark helicities $(\bar{h}) h$ in the case of longitudinal photons whereas for transverse photons there is an additional average over initial photon polarisations $\varepsilon$, and $f$ runs over the three light flavours. 
We use the photon wave functions $\psi^f_{h\bar{h},\varepsilon}(Q,z,\vec{r})$ in the form given in \cite{KMW},
and the GBW parametrisation \cite{GBW} for the dipole--proton cross section $\sigma(r) = \sigma_0\, N(r)$ with $N(r) = 1-\exp(-r^2/4R^2)$. The saturation radius in DDIS depends on $\xi$, $R(\xi)=(\xi/x_0)^{\lambda/2}$~GeV$^{-1}$
and $\sigma_0 = 23.03\; \mathrm{mb}$, $\lambda=0.288$, $x_0 = 3.04\cdot 10^{-4}$.

We performed the twist decomposition of Eq.\ (\ref{cross_sec_qq}) through the Taylor expansion in inverse powers of $QR$.

The contribution of the $q\bar q g$ component of $\gamma^*$ is calculated at $\beta = 0$ and in the soft gluon approximation (the longitudinal momentum carried by a gluon is much lower then carried by the $q\bar q$ pair). The correct $\beta$-dependence is then restored using a method described in \cite{Marquet}, with kinematically accurate calculations of Ref.~\cite{Wusthoff97}. The soft gluon approximation is valid in the crucial region of $M_X^2\gg Q^2$ or $\beta \ll 1$, where the deviations from DGLAP are observed. With these approximations one obtains, in consistence with \cite{Kov-Levin}:
\be
\label{cross_sec_qqg}
\frac{d\sigma^{q\bar{q}g}_{L,T}}{dM_X^2} = 
\frac{A}{M_X^2}
\sum_f\int d^2r_{01}
N^2_{q\bar{q}g} 
\sum_{\mathrm{spins}}\int_0^1 dz \,|\psi^f_{h\bar{h},\lambda}
|^2, 
\ee
\vspace*{-5ex}
\[
N^2_{q\bar{q}g} 
= \int d^2 r_{02}\, K_{01|2}\, (N_{02} + N_{12} - N_{01} - N_{02}N_{12})^2,
\]
where 
$A = N_c \sigma_0 ^2 \alpha_s \,/\, 32\pi^3 b_D$ and $N_{ij} = 
N(\vec{r}_j-\vec{r}_i)$, 
$\vec{r}_{01}, \vec{r}_{02},\vec{r}_{12}=\vec{r}_{02}-\vec{r}_{01}$ denote the relative positions of quark and antiquark $(01)$, quark and gluon $(02)$ in the transverse plain, and $K_{01|2} = r^2_{01}/r^2 _{02} r^2 _{12}$ is proportional to the dipole splitting kernel \cite{Mueller}. The form of $N^2_{qqg}$ follows from the Good-Walker picture of the diffractive dissociation of the photon \cite{Munier_Shoshi}. 
The factor $1/M_X^2$ is a remnant of the phase space integration under the soft gluon assumption. The twist decomposition of (\ref{cross_sec_qqg}) is performed using the Mellin transform in the $r_{01}$ variable:
\beq
\label{cross_sec_qqg_mellin}
\frac{d\sigma^{q\bar{q}g}_{L,T}}{dM_X^2} &=& 
\frac{A}{M_X^2}
\int\frac{ds}{2\pi i}\left(\frac{4 Q_0^2}{Q^2}\right)^{-s}\tilde{H}_{L,T} (-s) 
\, \tilde N_{qqg}^2(s),
\eeq
where the expression for $\tilde H_{L,T}(s) = \sum_f \tilde{H}_{L,T}^f(s)$ can be found in \cite{KMW}. The Mellin transform of $N^2_{q\bar{q}g} (r_{01})$ can be done in two steps. First one defines new integrals $\tilde{N}_{q\bar{q}g}^2(s) = I_1-I_2$ where, for $a=1,2$ one has
$I_a= (Q_0^{2s} / \pi) \int d^2r_{01} (r_{01}^2)^{s-1} \int d^2r_{02} \,K_{01|2}
\, S_a$
with $S_1 = (N_{02} + N_{12} - N_{02}N_{12})^2-(N_{01})^2$ and $S_2 = 2N_{01}(N_{02}+N_{12}-N_{02}N_{12}-N_{01})$. The integral $I_1$ can be performed exactly,
\beq
& & I_1=\pi \, (Q_0 R)^{2s} \, 2^{1+s}\,(2^{1+s}-1)\, \Gamma (s) \nonumber \\
& & \times \left[H_s - {}_3F_2(1,1,1-s;2,2;-1)s\right], \;\, H_s = \sum_{k=1}^s\frac{1}{k},
\eeq
and for $I_2$ we use the large daughter dipole approximation $r_{02} \gg r_{01}, \vec{r}_{12}\approx \vec{r}_{02}$ and obtain,
\beq
& &
I_2=\, \pi \, (Q_0 R)^{2s} \, 2^{1+2s} \, \Gamma (s) \, 
\left\{ 1-2^{1-s}+3^{-s} \right. \nonumber \\
& &
\left. + \, \frac{2^{-s}s}{1+s}\left[1 - {}_2 F_1\left( 1+s,1+s;2+s;-\frac{1}{2}\right)\right]\right\} . 
\eeq
These formulae agree within a few percent with exact numerical calculation of the diffractive cross-section~(\ref{cross_sec_qqg}).
The twist decomposition follows from (\ref{cross_sec_qqg_mellin}) as a sum over residues at the negative integer values of $s$. 

In the region of low $\xi$ the dominant contribution to the diffractive cross-section comes from $q \bar q g$.
However, at very low $\beta$, an even more resolved photon fluctuation, with two emitted gluons, $q\bar q g g$, may become relevant. Hence, at the lowest $\beta$, one expects some underestimation of the 
dipole model predictions with $q\bar q$ and $q\bar q g$ states only. Therefore we use the data with $\beta > 0.035$ in quantitative analysis. 
\\

\noindent
{\bf 4. The evidence of higher twists observation.}
The twist decomposition of the saturation model provides an efficient tool to study the HT effects in data. We focus on the region of $Q^2 \leq 6$~GeV$^2$, where the accuracy of DGLAP description breaks down. We compare the data to four different theory descriptions: (a) ZEUS-SJ NLO DGLAP fit \cite{ZEUS} to all data with $Q^2>5$~GeV$^2$, 
(b) the complete saturation model in the formulation of Refs.~\cite{Munier_Shoshi, Marquet} (MMS-Sat), (c) the GBW saturation model \cite{GBW}, and (d) the saturation models in which we retain only twist-2, twist-2 and 4 or twist-2, 4 and 6 contributions. In Fig.~2 we compare selected results with data: the extrapolated DGLAP results, DGLAP plus MSS-Sat twist-4, DGLAP plus twist-4 and twist-6 MMS-Sat, and the other results are only described in the text.

\begin{figure}

\centerline{
\includegraphics[width=1\columnwidth,height=0.65\columnwidth]{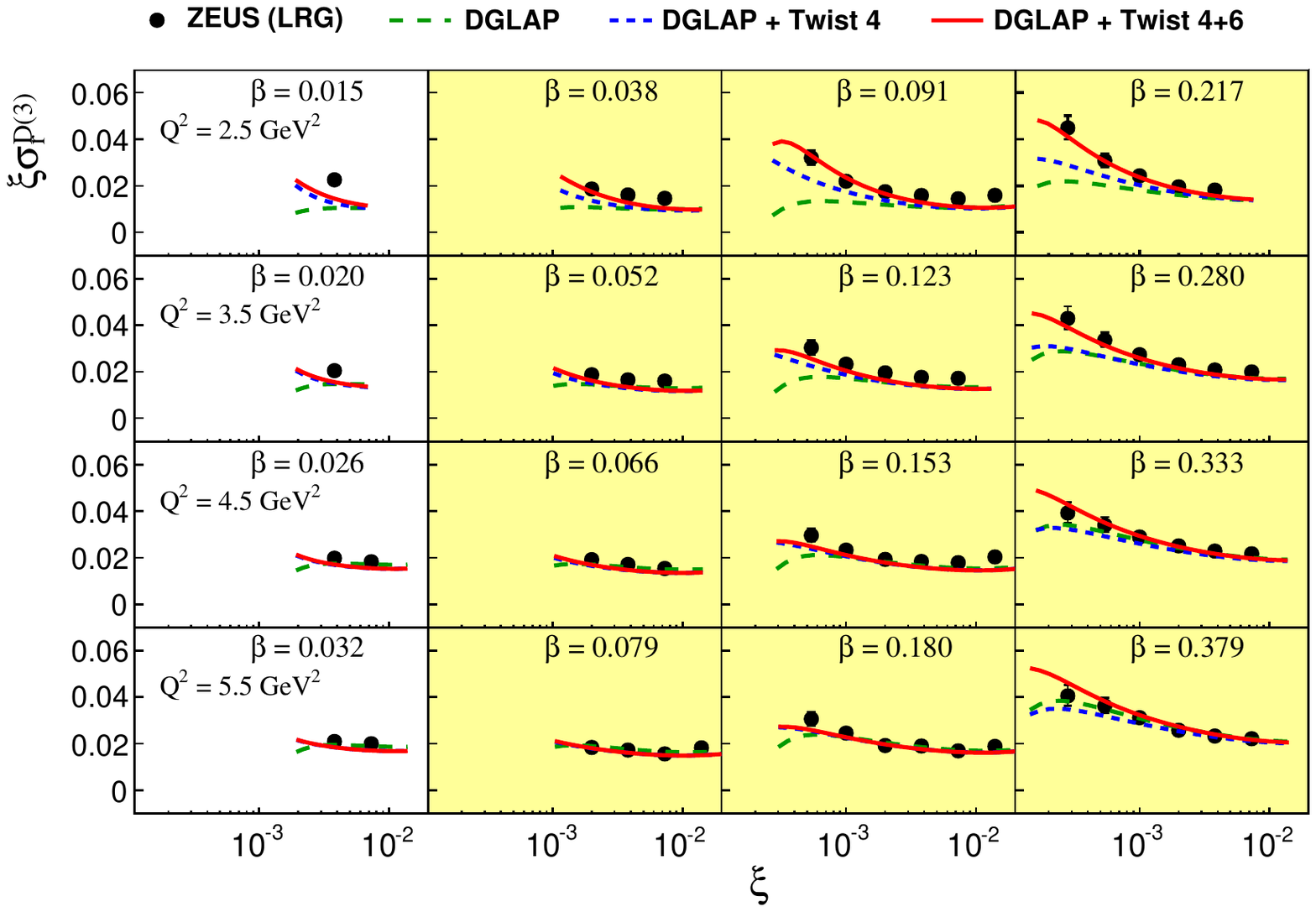}\vspace{-1em}
}

\caption{The LRG ZEUS data for $\xi\sigma_r ^{D(3)}$ at low $Q^2$ compared to a DGLAP fit \cite{ZEUS} and the DGLAP fit with included twist-4 and twist-4 and~6 corrections from the MMS saturation model. In yellow (gray) --- the region of $\beta$ where the correction due to $q\bar q gg$ may be neglected.
\label{fig2}
}
\end{figure}

The saturation model results are obtained including $q\bar q$ and $q\bar q g$ components, using the original GBW parameters $\lambda$ and $\sigma_0$, and three massless quark flavours. In our approach we modified the GBW parameter $x_0$ to $\xi_0 = 2 x_0$ in order to account for the difference between Bjorken~$x$ and $\xi$, the variables used as '$x$' in GBW dipole cross-section in DIS and DDIS respectively. The variable $\xi$ corresponds to the actual momentum fraction flowing from the proton to the photon, and Bjorken~$x$ in DIS is only the threshold value of~$x$ in hadroproduction. The actual $x$ of the gluon is larger, since typically, a hadronic mass $M^2 _h \sim Q^2$ is produced in DIS. We chose $\alpha_s = 0.4$ that provides a good description of data.

The conclusions from the analysis and from the figure are the following:
(i) at twist-2 the DGLAP fit extrapolation, and the twist-2 components of the MMS-Sat and GBW agree with one another, but all fail to describe the data below $Q^2 = 5$~GeV$^2$ and at low~$\xi$; the consistence of all approaches at twist-2 verifies reliability of saturation models at twist-2 (ii) the complete saturation models, MMS-Sat and GBW give similar predictions, better than the DGLAP description of the low~$Q^2$ data, but still the predictions are much below the data; (iii) the combination of twist-2 DGLAP and twist~4 part of MMS-Sat gives similar results as the complete MMS-Sat --- the data are not well described; (iv) a combination of the DGLAP fit and twist-4 and twist-6 components of MMS-Sat gives a good description of the data at low $Q^2$; (v) in GBW one finds only twist-2 and twist-4 contributions, both close to those of MMS-Sat, and GBW is unable to describe the data with any truncation of the twist series. We also found that the large HT effects in the saturation models appear only in the $q\bar qg$ component.     

The quality of the improved description of the data with our best model, NLO DGLAP + twist-4 and twist-6 components of MMS-Sat is quantified with $\chi^2 / \mathrm{d.o.f.}$ in Fig~1, the dashed curve. The ZEUS LRG data \cite{ZEUS} were fitted in this model for $Q^2 > Q^2 _{\mathrm{min}}$. In the fits only the DGLAP parameters were adjusted, and the twist-4 and~6 components were taken from the MMS-Sat model, with fixed parameters $\alpha_s = 0.4$, $\xi_0 = 2x_0$. Inclusion of HT terms greatly improves the fit quality in the low $Q^2$ region, --- the  maximal value of $\chi^2 / \mathrm{d.o.f.} \simeq 1.5$ at $Q^2_{\mathrm{min}} = 2$~GeV$^2$ is much lower than $\chi^2 / \mathrm{d.o.f.} \simeq 3$ of the DGLAP fit. The improved $\chi^2$ is not yet perfect, but it should be expected, as the MMS-Sat model provides only a first and crude estimate of HT effects in DDIS.    

The final conclusions follow: (1) the discrepancy between the twist-2 DGLAP description and the data has strong $Q^2$ and $\xi$-dependence, consistent with large contributions of twist-4 and twist-6; (2) the data require a truncation of the twist series in the saturation model. \\

\noindent
{\bf 5. Discussion.} We used an arbitrary truncation of higher terms in the twist expansion of the saturation models in order to determine a QCD motivated parametrisation of the $\xi$ and $Q^2$ dependence of the \hbox{twist-2} DGLAP breaking effects in DDIS data. A truncation of this kind, however, may be motivated in QCD. Let us recall that the GBW model assumes an eikonal scattering of the colour dipole off proton. This means, that one couples to the dipole an arbitrary number  $t$-channel gluons, without significant suppression. This is, however, in conflict with results of the BFKL framework (see e.g.\ Ref.~\cite{BFKL}). In BFKL, at the leading logarithmic approximation, only one reggeized gluon may couple to a fundamental colour line. Since DGLAP and BFKL approximations have the same double logarithmic ($\ln x \,\ln Q^2$) limit, one concludes that also in DGLAP couplings of more than two gluons to a colour dipole is much weaker than in the eikonal picture. This, in turn, has implications for the HT representation in the saturation model, because a $n$-gluon exchange provides the leading \hbox{twist-$n$} contribution in DIS cross-sections. Thus, we conclude that the eikonal saturation model should overestimate all the HT terms in single colour dipole scattering.    

In DDIS, however, also the $q\bar qg$ component is important, equivalent to two colour dipoles in the planar limit. Thus, one may couple up to four gluons (two per dipole) to the scattering state, without the BFKL constraint of one-per-line reggeized gluon coupling. Such a coupling, at the amplitude level, means that one may expect unsuppressed contributions up to twist-8 in the cross-section, and a suppression beyond twist-8. This qualitative argument provides possible  motivation for truncations of the twist series in the saturation model, but in fact, in order to obtain reliable results, multi-gluon couplings to a $q\bar q g$ system need to be computed in QCD.

Finally, let us point out that so far the HT terms in the saturation model were too weak to be seen in data, so a truncation of HT terms in saturation models should not lead to deterioration of other data description.\\

{\small \paragraph{Acknowledgements.} 
The work is supported by the Polish National Science Centre grant no.\ DEC-2011/01/B/ST2/03643.}

\end{document}